# Distributed fiber sparse-wideband vibration sensing by sub-Nyquist additive random sampling

Jingdong Zhang,[1][†] Hua Zheng,[1] Tao Zhu,[1,*] Guolu Yin[1], Min Liu,[1] Yongzhong Bai,[2] Dingrong Qu,[2] Feng Qiu,[2] And Xianbing Huang[2]

[1]Key Laboratory of Optoelectronic Technology & Systems (Ministry of Education), Chongqing University, Chongqing 400044, China
[2]State Key Laboratory of Safety and Control for Chemicals, SINOPEC Research Institute of Safety Engineering, Qingdao 266000, China
*Corresponding author: zhutao@cqu.edu.cn
†: zjd@cqu.edu.cn

**The round trip time of the light pulse limits the maximum detectable vibration frequency response range of phase-sensitive optical time domain reflectometry (φ-OTDR). Unlike the uniform laser pulse interval in conventional φ-OTDR, we randomly modulate the pulse interval, so that an equivalent sub-Nyquist additive random sampling (sNARS) is realized for every sensing point of the long interrogation fiber. For an φ-OTDR system with 10 km sensing length, the sNARS method is optimized by theoretical analysis and Monte Carlo simulation, and the experimental results verify that a wide-band spars signal can be identified and reconstructed. Such a method can broaden the vibration frequency response range of φ-OTDR, which is of great significance in sparse-wideband-frequency vibration signal detection, such as rail track monitoring and metal defect detection.**

xxxx

Distributed vibration sensing by phase-sensitive optical time domain reflectometry (φ-OTDR) has been intensively studied for its potential applications in fields like intrusion detection, oil and gas pipeline monitoring, high speed rail or urban rail transit monitoring, and so on [1-5]. It is important for the sensing distance and the frequency response range to meet the requirements of practical applications. Up to now, ultra-long φ-OTDR with 175 km sensing range has been achieved [5], but its vibration frequency response range is limited to less than 285 Hz. In order to broaden the frequency response range, two kinds of approaches have been proposed: the combination of interferometer with φ-OTDR [2], and the frequency multiplexing method [3,4].

Taking the φ-OTDR based high-speed rail track monitoring system as an example, the length of sensing fiber should be larger than 50 km or even 100 km, which corresponds to a maximum detectable frequency of 1 kHz or 500 Hz, respectively. However, in terms of the wave propagation along a railway track, the frequency of train-induced vibration is up to 80 kHz [6]. It is difficult for the conventional φ-OTDR system to detect such a broadband signal.

In some cases, the vibration waves are sparse in frequency domain, which gives a chance for φ-OTDR system to handle with low pulse repetition rate. As for rail track monitoring, most of the long-range waves decay rapidly because only several vibration modes with certain frequencies are supported [7], owing to the fact that the vibration waves degenerate to sparse signal [8] over 10 kHz to 60 kHz. It should be noted that sparse signal is also used widely in many fields, such as metal damage detection [9].

In this letter, we demonstrate a new φ-OTDR system inspired by the concept of non-uniform sampling [10-12]. Thanks to its anti-aliasing characteristic in frequency domain, the proposed method, named sub-Nyquist additive random sampling (sNARS) technique, is capable of sampling the wide-band sparse frequency signals with sub-Nyquist sampling rate [8, 12]. The sNARS method is studied theoretically and verified with Monte Carlo simulation, by which the key parameters for φ-OTDR vibration sensing are optimized. Then a sNARS based φ-OTDR system is established by randomly modulating the laser pulse interval. The sparse signals, generated by driving the piezoelectric transducer (PZT) with large modulation depth $K$ ($K>π$), are sampled and reconstructed with a dynamic range over 30 dB. The proposed system exhibits a broadband response for sparse vibration signal sensing without broadening the bandwidth of detection, which provides a potential solution to monitor rail track and detect metal damage.

Generally, the signal $x(t)$ sampled by the sampling pulse $s(t)$ can be written as $y(t)=x(t)s(t)w(t)$, where $s(t)$ is defined as $s(t)=\sum_{n=-\infty}^{\infty}\delta(t-t_n)$, $w(t)$ is the rectangle window function whose width is the total sampling time. Then the power spectral density (PSD) of $y(t)$ is $W_y(f)=W_x(f)*W_s(f)*W_w(f)$, where $*$ denotes convolution, and $W_x(f)$, $W_s(f)$, $W_w(f)$ are the PSDs of $x(t)$, $s(t)$ and $w(t)$, respectively. In the uniform sampling with $t_n=n/f_s$, $W_s(f)$ is a Dirac comb function with period $f_s$ [13] as shown in Fig. 1(a). Therefore, $W_y(f)$ is the periodic replicas of $W_x(f)*W_w(f)$ with period $f_s$. The under test signal $x(t)$ must be bandlimited within $f_s/2$ to avoid frequency aliasing.

Unlike the uniform sampling case, the interval $\Delta t_n=t_n-t_{n-1}$ of additive random sampling (ARS) is an independent identically distributed (IID) random variable with probability density function (PDF) $p(t)$.

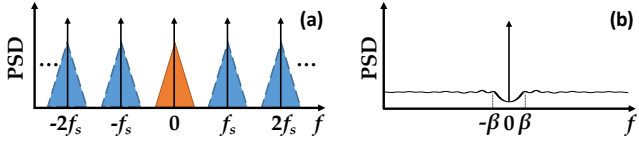

**Fig. 1.** Power spectral density of uniform sampling pulse (a) and additive random sampling pulse (b).

Shapiro [10] and Beutler [11] established that the analytic expression of $W_s(f)$ of ARS is

$$W_s(f) = \begin{cases} \beta \operatorname{Re}\left\{\dfrac{1+\phi(f)}{1-\phi(f)}\right\}, & f \neq 0 \\ \beta^2 \delta(f), & f = 0 \end{cases}, \quad (1)$$

where $\beta=E(1/\Delta t_n)$ is the expectation of the sampling rate and $\phi(f)=\int_{-\infty}^{+\infty} p(t)e^{-j2\pi f\cdot t}dt$ is the characteristic function of $p(t)$. As shown in Fig. 1(b), $W_s(f)$ contains a delta function at the origin, replicating $W_x(f)*W_w(f)$ via the convolution operation, and a wideband noise component with amplitude much smaller than the original spectrum when $\beta \gg 1$. In this way, the aliases are suppressed and the mean sampling rate $\beta$ no longer needs to be larger than the Nyquist frequency. For the actual sampling data $x(t_n)$ with finite sampling length $N$, the PSD spectrum of the signal can be evaluated by $W(f)=N|X(f)|^2$, in which $X(f)$ is the non-uniform discrete Fourier transform (NDFT) of $x(t_n)$ as

$$X(f) = \dfrac{1}{N}\sum_{n=1}^{N} x(t_n) e^{-j2\pi f\cdot t_n}. \quad (2)$$

With Eq. (2), the information of the under test signal, including the frequencies, amplitudes and phases, can be identified.

In order to customize an sNARS scheme for the sparse vibration signal sensing in φ-OTDR system, the specific parameters of sampling should be optimized. Considering the fact that the minimum pulse interval in φ-OTDR system must be longer than the laser pulse round trip time $t_r=2ln_{eff}/c$, the sampling interval $\Delta t_n$ is set to be a random variable with a uniform distribution $\Delta t_n \sim U(\Delta t_{min}, \Delta t_{max})$, where $\Delta t_{min}=t_r$, $l$ and $n_{eff}$ are the length and the effective index of the sensing fiber, and $c$ is the speed of light in vacuum. Assuming $\Delta t_{max}=k_t\Delta t_{min}$, the sampling interval factor $k_t$ affects the performance of the sampling. Besides, when generating the random sampling trigger pulses with equipment such as arbitrary waveform generator (AWG), $\Delta t_n$ is quantized onto a fixed time grid determined by the sampling rate of the AWG $f_{s-awg}$, which also affects the NDFT results. It should be pointed out that owing to the processing gain of discrete Fourier transform and the truncation effect [13], the sampling length $N$ and analytical frequency resolution $f_a=\beta/Nk_a$ affect the signal to noise ratio (SNR) and the measured amplitude of the under test signal. These aforementioned key parameters are analyzed theoretically and verified by Monte Carlo simulation.

The alias suppression effect of sNARS is demonstrated firstly in sampling of a single frequency signal $x(t)=A\sin(2\pi ft)$, where $A=1$ and $f=100\pi$ kHz. The minimum interval $\Delta t_{min}$ is set to 100 μs, which corresponds to 10 km sensing fiber. By taking the NDFT over 0~500 kHz, the PSD spectrum of the sNARS is compared with its theoretical value from Eq. (1). All the PSD spectra are normalized with their max values. Figure 2(a) and Visualization 1 show the PSD spectra with different $k_t$, in which the SNR is defined as $SNR=10\log(W_a/W_n)$ where $W_a$ is the measured PSD of the under test signal, and $W_n$ is calculated by the root-mean-square (RMS) of the noise amplitude. $W_n$ and theoretical value are marked by green dotted line and orange dashed line, respectively. When $k_t$ approaches 1, the aliases raise while the sampling degrades into a uniform one. With the increase of $k_t$, the frequency aliasing peaks decrease rapidly and form a flat noise floor which coincides with the green dotted line. Considering the total sampling time, $k_t=2$ is adopted in the following experiments. As illustrated in Fig. 2(b) and Visualization 2, the frequency aliasing effect is eliminated while the $f_{s-awg}$ exceeds 1000 kHz, which is because the sNARS with quantized sampling interval can be considered as the re-sampling of a uniform sampling data with sampling rate $f_{s-awg}$. In a nutshell, the anti-aliasing effect of the proposed sNARS method is determined by the $k_t$ and $f_{s-awg}$, and appropriate values should be adopted when the trigger intervals of the laser pulses are generated.

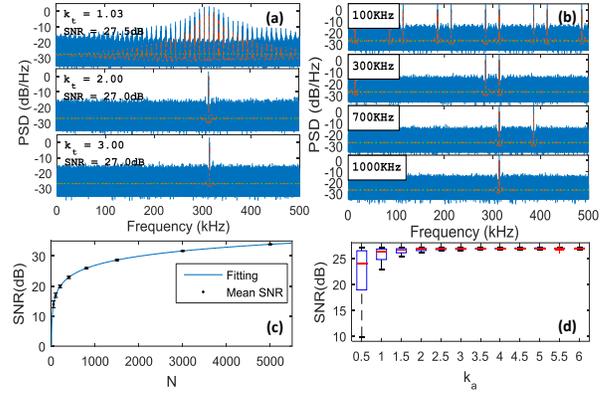

**Fig. 2.** Theoretical analysis and Monte Carlo simulation of sampling parameters. (a) PSD spectra with different maximum sampling interval factor $k_t$. (b) PSD spectra with different AWG sampling rate $f_{s-awg}$. (c) Average SNR with different sampling length N. (d) Boxplot graph of SNR with different analytical frequency resolution factor $k_a$. Visualization 1-4 (in Supplementary Material) are the overall PSD spectra of (a)-(d) with all parameter settings.

To test the sampling length $N$ and analytical frequency resolution factor $k_a$, 50 signals with various frequencies $f$ and same amplitude $A=1$ are sampled. Visualization 3 displays the PSD spectra with separate sampling length $N$. Fig. 2(c) shows the average SNR has a logarithmic correlation with the increasing of $N$ due to the processing gain [13]. Visualization 4 demonstrates the PSD spectra with different $k_a$. The PSD peaks and SNR fluctuate while $k_a<2$ because $f_a$ is nearly the same value with the sNARS frequency resolution $f_r=\beta/N$. By using a boxplot graph, shown in Fig. 2(d), the SNR is gathered statistically, showing that the SNR is stable when $k_a>4$. Hence, $k_a=4$ is used in the following experiment.

After obtaining the optimum parameters, we evaluate the proposed sNARS method in a direct detection φ-OTDR vibration sensing system by comparing it with the conventional Nyquist uniform sampling (NUS). The experimental setup is shown in Fig. 3(a), in which the narrow linewidth laser source (NKT Laser, E15) is modulated into random interval pulses by the acoustic-optic modulator (AOM, Gooch & Housege, Fiber-Q) driven by an AWG (Tektronix, AWG5012C). 2 m fiber is wrapped on the PZT tube. In this direct detection φ-OTDR system, the intensity response to the external phase modulation $\theta(t,z)$ can be written in the following form [14]

$$I(t,z) = D(z) + A(z)\cos[\theta(t,z) + \phi(z)], \quad (3)$$

where the direct-current component $D(z)$, the alternating current component $A(z)$ and the initial phase $\phi(z)$ are slowly varying with time. If a single frequency signal $x(t) = A_v\cos(2\pi ft)$ is applied to the PZT, the intensity response of the fiber at the place corresponding to the PZT is

$$I_z(t) = D_z + A_z\cos[A_v K_v \cos(2\pi ft) + \phi_z], \quad (4)$$

where $A_v$ is the amplitude of driving voltage, $K_v$ is the phase-voltage response coefficient of PZT. According to the Jacobi–Anger expansion, $I_z(t)$ can be written in the sum of harmonic waves as

$$I_z(t) = D_z + A_z\cos(\phi_z)[J_0(K) + 2\sum_{n=1}^{\infty}(-1)^n J_{2n}(K)\cos(4\pi nft)] \\ + 2A_z\sin(\phi_z)\sum_{n=1}^{\infty}(-1)^n J_{2n-1}(K)\cos[2\pi(2n-1)ft], \quad (5)$$

where $J_n(K)$ is the $n$-th Bessel function of the first kind and $K=A_v K_v$ is the modulation depth. When $K \ll \pi$, the high order Bessel function values are negligible compared with first order Bessel function value and hence the intensity response is a single-frequency signal with the same frequency of the driving voltage. Comparatively, when the modulation depth $K$ is larger than $\pi$, i.e. the driving voltage is large or the driving frequency is near the resonant frequency of the PZT, the harmonic frequency components will show up. At this point, the intensity response of the φ-OTDR system is considered to be a sparse frequency signal [8, 12], and we employ this signal to test the property of the proposed sNARS method in the φ-OTDR system.

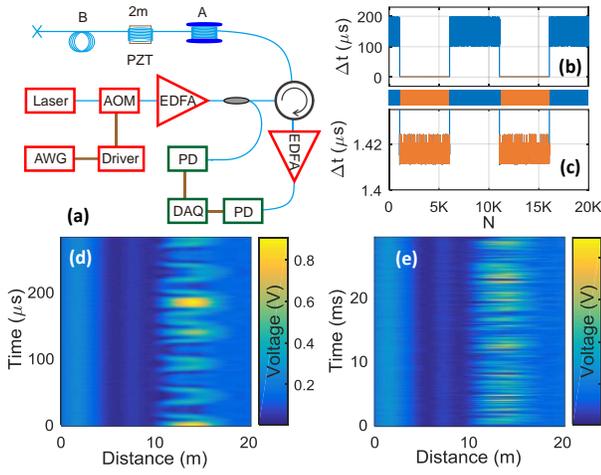

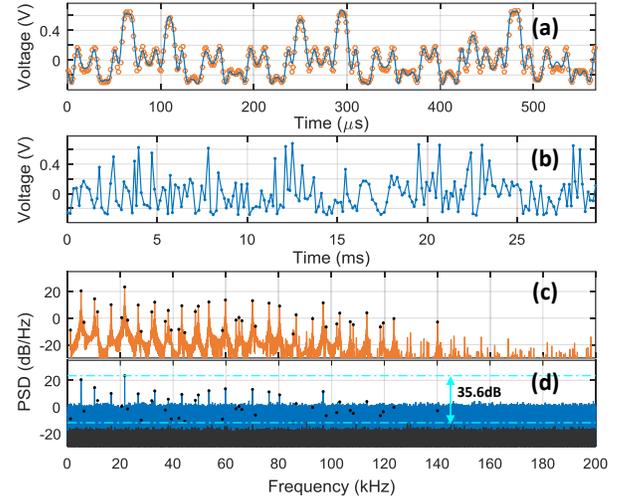

**Fig. 3.** (a) The experimental setup. (b-c) The sampling interval of NUS and sNARS. (d) Nyquist uniform sampling traces. (e) Sub-Nyquist additive random sampling traces.

**Fig. 4.** (a) Vibration sampling data of NUS (orange circles) and reconstructed signal from sNARS (blue line). (b) Vibration sampling data of sNARS. (c) PSD spectrum of NUS sampling data. (d) PSD spectrum of sNARS sampling data. Visualization 5 (in Supplementary Material). The signal reconstruction process.

In order to sample the vibration point with high sampling rate, we employ short sensing fibers A and B with length of 15 and 35 m, respectively, as shown in Fig. 3(a). By modulating the interval of the laser pulse, for a certain point along the sensing fiber, the NUS (sampling interval $\Delta t_u$=1.4164 μs, sampling length $N_u$=5000) and the sNARS (sampling interval $\Delta t_r$ obeys uniform distribution $U$(100 μs, 200 μs), sampling length $N_r$=5000, $f_{s\text{-}awg}$=20 MHz) are implemented alternately. The full width at half maximum (FWHM) of the laser pulse is 50 ns. Since $\phi(z)$ is slowly varying with time, the vibration signal keeps unchanged during different sampling periods and the proposed sNARS method can be compared with the conventional NUS method. It should be noted that the sampling rate for every backscattering trace is uniform (2 GSa/s), and one backscattering trace provides one sampling data of the vibration point. Figure 3(b) shows the sampling interval of NUS and sNARS ($\Delta t_u$ and $\Delta t_r$, respectively), which are obtained from the triggered timestamp data of the data acquisition card (DAQ, Gage, CSE24G8). $\Delta t$ axis is enlarged in Fig. 3(c), in which the time ~10 ns jitter is normal for uniform sampling and can be neglected. The color bar between Fig. 3(b) and (c) illustrates the NUS period (orange) and sNARS period (blue). When a driving signal containing two frequency components (6 Vpp, 5.13 kHz and 1.6 Vpp, 21.634 kHz) is applied to the PZT, the first 200 of the 5000 sampling traces of NUS and sNARS are illustrated in Fig. 3(d) and (e), respectively. It is obvious that the Fig. 3(d) displays the detail of the vibration signal, while Fig. 3(e) exhibits the vibration signal at random instants.

The sampling data of the vibration points of NUS and sNARS, shown in Fig. 4(a) and (b), are analyzed and compared in frequency domain and time domain. Figure 4(c) shows the PSD of the NUS data acquired by fast Fourier transformation (FFT), and Fig. 4(d) exhibits the PSD of the sNARS data acquired by NDFT. Despite the fact that the alias noise floor emerges in Fig. 4(d), the frequency components of the signal can still be identified by signal reconstruction. The reconstruction process, inspired by matching pursuit [15], is shown in Visualization 5. The frequency components are identified iteratively in order of their amplitudes, by which the alias noises of the frequency components with large amplitudes will be eliminated. To accurately estimate the frequencies $f_i$ and the amplitudes $A_i$, Lorentzian fittings are taken within the main lobes of the Sinc function shaped peaks of every frequency component at each iteration. The phase $\varphi_i$ can be obtained by calculating the phase angle of NDFT at $f_i$. Then the signal residual $x_r^i(t_n) = x_r^{i-1}(t_n) - A_i\cos(2\pi f_i t_n + \varphi_i)$ is used as signal data in the next iteration. The iterations are terminated when $A_i<3Q_{0.99}$, where $Q_{0.99}$ is the 0.99 quantile of the PSD of the signal residual. The PSD of signal residual at the last iteration, which can be regarded as the noise, is showed as gray line in Fig. 4(d). The $f_i$, and $A_i$ are marked as black dots in Fig. 4(c) and (d), in which the maximum signal component is 35.6 dB larger than the minimum signal component. With the parameters $f_i$, $A_i$ and $\varphi_i$ of every frequency component, the sparse signal can be reconstructed in time domain. In Fig. 4(a), the reconstructed signal is displayed by blue line. The reconstructed signal corresponds closely to the signal sampled at the high frequency sampling rate, which indicates that the proposed sNARS method can be used in φ-OTDR system for high frequency sparse signal detection and reconstruction with low sampling rate.

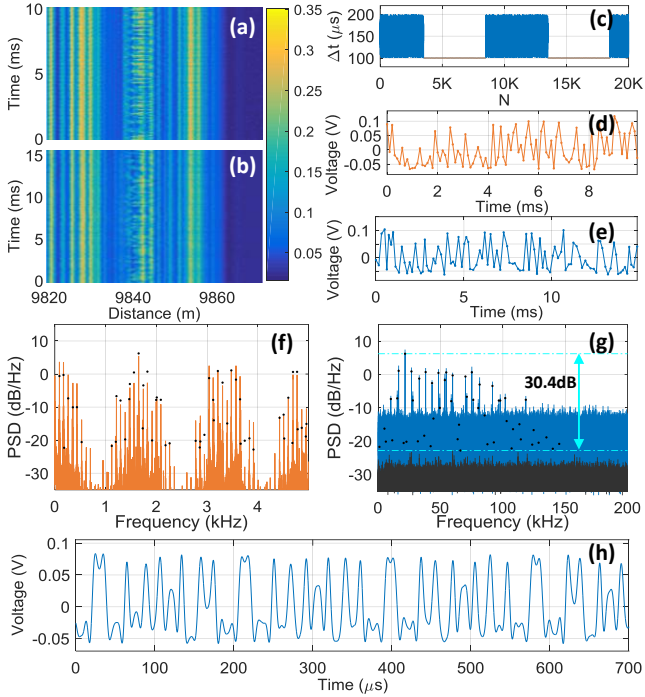

**Fig. 5.** (a) Sampling traces of NUS. (b) Sampling traces of sNARS. (c) The sampling interval of NUS and sNARS. (d) Vibration sampling data of NUS. (e) Vibration sampling data of sNARS. (f) PSD spectrum of NUS sampling data. (g) PSD spectrum of sNARS sampling data. (h) Reconstructed signal. Visualization 6 (in Supplementary Material). The signal reconstruction process.

Then the proposed sNARS and NUS are compared in a long-range φ-OTDR vibration sensing system, with A and B in Fig. 3(a) being 9836 and 35 m, respectively. The NUS (sampling interval $\Delta t_u=1.4164$ μs, sampling length $N_u=5000$) and the sNARS (sampling parameters are same with the ones in Fig. 3(b)) are also implemented by turns and the pulse interval $\Delta t$ of NUS and sNARS are plotted in Fig. 5(c) as orange and blue line, respectively. When driving the PZT with the same signal shown in Fig. 3, Fig. 5(a) and (b) display the first 100 of the 5000 period traces recorded traces of NUS and sNARS at the end section of the sensing fiber. The vibration sampling data of NUS. and sNARS are shown in Fig. 5(d) and (e), respectively. Taking FFT and NDFT of the corresponding 5000 points NUS and sNARS sampling data, the PSD spectra are shown in Fig. 5(f) and (g), respectively. The frequency components of the wide band sparse signal are aliased into 0 to 5 kHz in Fig. 5(f) because the signal bandwidth exceeds the Nyquist frequency of NUS. However, this signal can be identified correctly in Fig. 5(g). Visualization 6 shows the reconstruction process from the sNARS data, and the identified $f_i$, and $A_i$ are marked as black dots in Fig. 5(g). The PSD of signal residual at the last iteration is showed as gray line in Fig

. 5(g). 53 frequency components are identified and the maximum signal component is 30.4 dB larger than the minimum signal component. The alias frequencies of the reconstructed frequency components are calculated and marked as black dots in Fig. 5(f), which corresponds well with the peaks. The amplitude difference between the dots and peaks are mainly due to the change of $\phi(z)$. The time domain reconstructed signal is presented in Fig. 5(h). The comparison above reveals that the proposed sNARS based φ-OTDR system is aliasing proof and able to detect the sparse-wideband vibration signals in long sensing range.

In conclusion, a sNARS based long range φ-OTDR vibration sensing system is proposed, in which wideband sparse vibration signal with sparsity exceeding 50 can be identified and reconstructed. Since the redundancy is minimized compared with uniform sampling, the sNARS method has the ability of enhancing the frequency response without broadening the detection bandwidth, which can serve as a valuable complement for conventional φ-OTDR system.

**Funding.** Ministry of science and technology (2016YFC0801202); National Natural Science Foundation of China (61635004, 61475029, 61775023, 61405020); Science fund for distinguished young scholars of Chongqing (CSTC2014JCYJJQ40002).

**Acknowledgment.** The authors are particularly grateful to Leilei Shi, Yongfeng Liu and Iroegbu Paul Ikechukwu for their supports on this work.